\documentclass{elsart}
\usepackage{graphicx}
\usepackage{dcolumn,color}
\usepackage{bm,rotate,rotating}
\usepackage{epsfig}
\usepackage[english]{babel}

\begin{document}                                                               
\bibliographystyle{try}
\topmargin 0.1cm
\vspace*{10mm}
\begin{center}
{\large\bf HADRON SPECTROSCOPY WITHOUT \\
\vskip 3mm
CONSTITUENT GLUE
}

\vspace*{10mm}
{\sc Eberhard Klempt}

\vspace*{2mm}
{\it Helmholtz-Institut f\"ur Strahlen-- und Kernphysik \\
	Universit\"at Bonn\\
	Nu{\ss}allee 14-16, D-53115 Bonn, GERMANY\\
	\texttt{e-mail: klempt@hiskp.uni-bonn.de}
        }
\end{center}
\vspace*{20mm}
Abstract\\
Glueballs and hybrids are predicted to exist 
but searches for them have failed to provide conclusive
evidence. One--gluon exchange is not an important part of strong
interactions in this energy regime. Instead, quarks seem to interact
indirectly, via changes of the QCD vacuum. Strong interactions seem
to be governed by instanton--induced interactions; the chiral
soliton model gives a more suitable interpretation of the
$\Theta^+(1540)$ than models based on the dynamics of four quarks and
one antiquark.
\vfill
\begin{center}
Contributed to \\
10th International Symposium on
Meson-Nucleon Physics and \\
the Structure of the Nucleon\\
August 29 - Sept 4, 2004  \\
Beijing, China\\
\end{center}
\vspace*{10mm}
\clearpage

\section{The scientific scope}	

There is only little understanding of the dynamics of quarks 
and gluons in the intermediate energy regime where meson and baryon 
resonances dominate strong interactions. Chiral symmetry,
expected to hold for nearly massless quarks, is spontaneously broken, 
and quarks acquire an effective 'constituent' mass. In quark
models, mesons and baryons are thus described by constituent quarks in a
confining potential. The interaction between 
quarks is more complicated than just a confinement potential
suggests: the full interaction is parameterized by some kind 
of additional `residual' interaction, by `effective' one--gluon 
exchange, by exchange of pseudoscalar
mesons (i.e. Goldstone bosons), or by instanton induced interactions.
From deep inelastic scattering it is known that baryons are more complex.
The structure functions reveal a rich dynamical sea of
quark--antiquark pairs, but there is no bridge from the high--energy partonic
structure to the dynamics of 
constituent quarks and their interaction.
In recent years, two interpretations have been developed 
of strong interactions physics in the confinement region. 
One interpretation underlines the importance of the gluon fields. The
residual interaction between quarks is given by an effective 
one--gluon exchange, gluons can - like quarks - develop an effective
mass. Gluons manifest themselves in new degrees of freedom in spectroscopy,  
in glueballs and in hybrids. The proponents of this picture interprete
the $\Theta^+(1540)$ as pentaquark, as bound state of four quarks and
one (strange) antiquark. 
The  second view is proposed in the chiral 
soliton picture. Quarks interact dominantly by changing the
vacuum, like Cooper pairs interact via phonon exchange. 
In this picture, the $\Theta^+(1540)$ 
arises naturally as member of an anti-decuplet. Quarks interact
via instanton--induced interactions. 
The forces are transmitted by vacuum fluctuations
of the gluon fields, not as direct quark--quark interactions. 
Glueballs and hybrids are no obvious features in
this kind of theory. A recent experimental survey can be found
in~\cite{Klempt:2004yz}. 
\section{Gluon exchange or instanton--induced interactions 
in baryons\,?}
\vspace*{-5mm}
The three--quark valence structure of baryons supports a rich spectrum
which is very well suited to study the effective interactions between
quarks in resonances. 
Fig. 1 shows a Regge trajectory of $\Delta^*$ and of N$^*$
resonances having intrinsic spin $3/2$. \\ 
\begin{figure}[h!]
\begin{center}
\epsfig{file=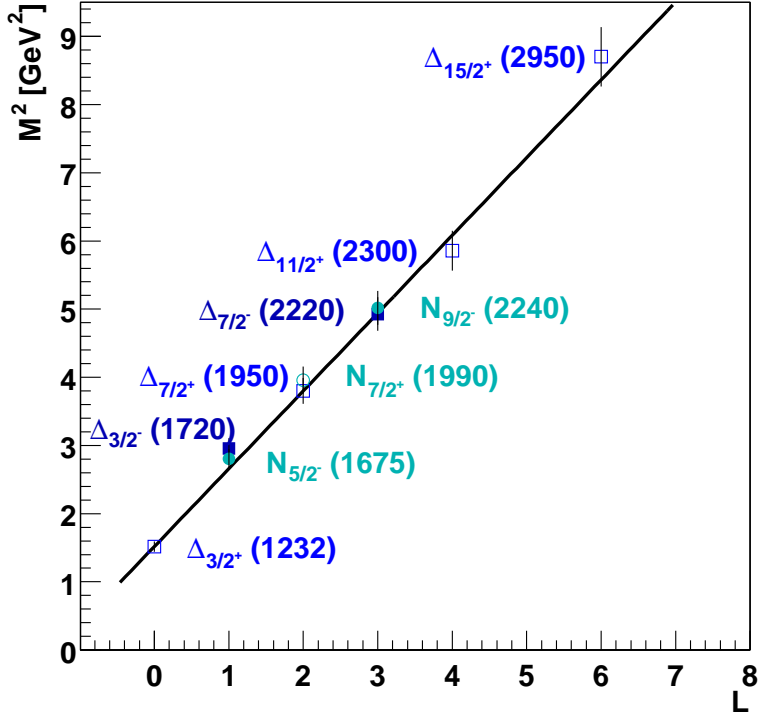,width=0.8\textwidth}
\end{center}
\caption{
The lowest--mass $\Delta^*$ resonances lie on Regge trajectories.  
If plotted against the intrinsic orbital angular momentum, 
also negative--parity resonances fall onto the trajectory. For even
parity the mass for $J=L+3/2$ is plotted, for odd parity that for
$J=L+1/2$. States with given $L$ but different $J$ are mass
approximately degenerate. This is the well known spin--orbit  
puzzle: from one--gluon exchange, large spin--orbit splittings are
expected. Surprising, perhaps, is the observation that nucleon
resonances with intrinsic spin $S=3/2$ are degenerate in mass with
the $\Delta$ series.
}
\end{figure}
\noindent
Nucleon resonances with intrinsic spin $1/2$ are discussed next. 
These can be separated into groups of states with 
even parity coming from a symmetric 56-plet; odd--parity baryons may
come from a 70-plet with mixed symmetry, from the totally
antisymmetric singlet system, or from a decuplet, see Fig. 2. 
The common feature
of each group is the fraction in the wave function which is
antisymmetric in spin and in flavor. This fraction is largest for
singlet baryons, reduced for
octet baryons from a 56-plet, even smaller for octet baryons from a
70-plet, and vanishes for decuplet baryons. For each group, there
seems to be a common shift in mass square. This shift is proportional
to the fraction 
of the baryon wave function which is antisymmetric in spin and in
flavour. This is a very characteristic pattern which must reflect the 
symmetry properties of the underlying interaction. Indeed,
instanton--induced interactions follow this symmetry. Thus the pattern
observed in Fig.~2 provides strong support for 
instanton--induced interactions being the residual interaction which
complements the confinement forces. The pattern can be formulated as
simple baryon mass formula having four parameters only. It 
reproduces very well the observed baryon mass spectrum, with a
$\chi^2$ which is much better than for a model based on one--gluon
exchange interactions (which suppresses spin--orbit effects by
arbritrarily assuming that spin--orbit forces and the Thomas precession 
in the confinement field compensate each other). 
\begin{figure}
\epsfig{file=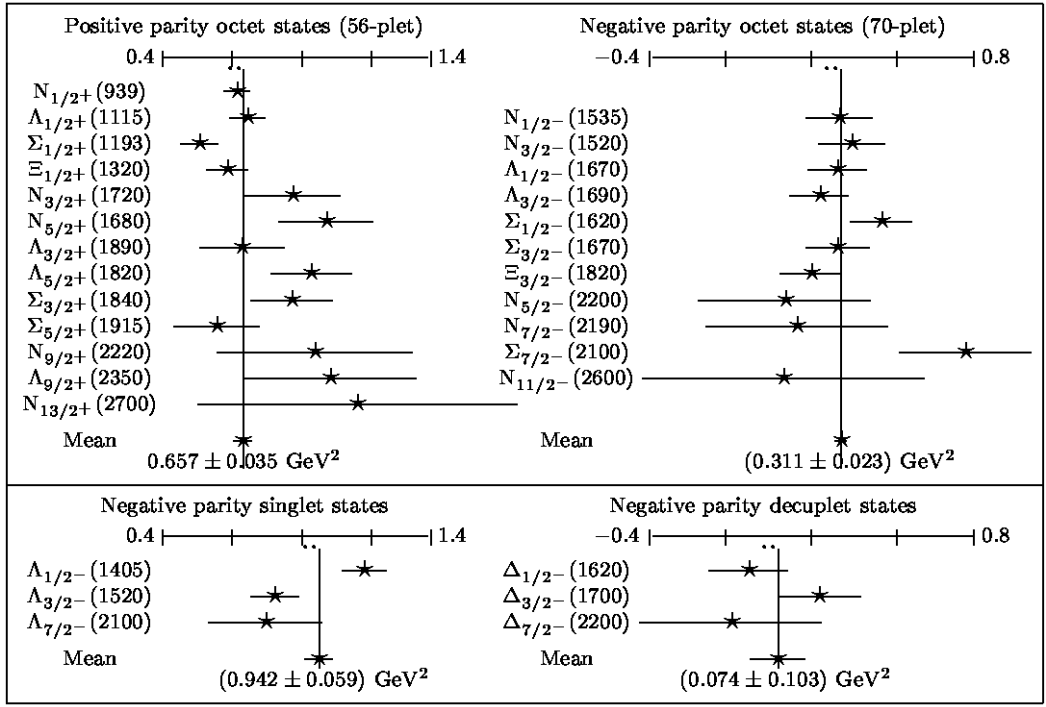,width=1\textwidth}
\caption{Mass shift (in GeV$^2$) with respect to the $\Delta$ Regge
trajectory. The nucleon has a (squared) mass of 0.88\,GeV$^2$, the 
$\Delta(1232)$ of 1.52\,GeV$^2$. The difference, 0.64\,GeV$^2$, is 
plotted. For resonances with strangeness, the Regge
trajectory starts at the $\Sigma^*(1385)$ mass but has the same slope.}
\end{figure}
\vspace*{4mm}
\section{Is there convincing evidence for glueballs\,?}
Glueballs, hybrid mesons and hybrid baryon are predicted
by QCD inspired models and may even be a consequence of QCD on the
lattice. But inspite of intensive searches,
no convincing evidence for their discovery has been reported. 
\vspace*{-6mm}
\subsection{Search for the pseudoscalar glueball}
\vspace*{-8mm}
The Particle Data Group\cite{Eidelman:2004wy} decided in their 2004 edition
that there is sufficient evidence that the former $\eta(1440)$ is
split into two components, the $\eta(1405)$ component decaying mostly
into $a_0(980)\pi$ and $\eta\sigma$,  
and the $\eta(1474)$ with $\rm K{}^*\bar K$ as preferred decay mode.
The following interpretation of the pseudoscalar mesons
is offered:
\vspace*{-2mm}

\begin{center}
\hspace*{6mm}\fbox{\parbox{90mm}{
\vspace*{-2mm}
\begin{center}
$\pi$\qquad\quad\quad
$\eta$\qquad\qquad\qquad\quad\quad
$\eta^{\prime}$\qquad\quad  $K$ \\
{$\pi(1300)$\quad\
$\eta(1295)$}\quad\ {$\eta(1405)$}\quad\
$\eta(1475)$\ $K(1460)$ \\
${n\bar n}$\qquad\quad {${n\bar n}$}\qquad\ 
{glueball}\qquad\ 
{${s\bar s}$}\qquad\ ${n\bar
s}$\qquad\quad  \\ 
\end{center}
\vspace*{-2mm}
}}
\end{center}

The $\eta(1295)$ and $\pi(1300)$ mesons have the same masses,
hence the $\eta(1295)$ and $\eta(1475)$ have likely
a nearly ideal mixing angle, with $\eta(1475)$
being the $s\bar s$ state. The $\eta(1405)$ does not find a slot in 
sthe spectrum of $\bar qq$ mesons; 
the low mass part of the $\eta (1440)$ could be a glueball.
\par
Radiative J/$\psi$ decays show an asymmetric peak in the $\eta(1440)$
region. Both $\eta(1405)$ and $\eta(1475)$ contribute to
the process. Radial exciations are hence produced in radiative 
J/$\psi$ decays (not only glueballs). But then, $\eta(1295)$ should
also be produced, but it is not\,! There is also no evidence for 
$\eta(1295)$ from $\gamma\gamma$ fusion at LEP, nor from  a study of
$\rm\gamma\gamma\to K^0_sK^{\pm}\pi^{\mp}$, but
$\eta(1440)$ is seen\cite{Acciarri:2000ev}.
The $\eta$(1440) coupling 
to photons is much stronger than that of $\eta$(1295): 
the assumption that the $\eta$(1295) is a $(u\bar u+d\bar d)$
radial excitation must be wrong\,!
The mass of the pseudoscalar resonance in
$\gamma\gamma$ fusion is about 1460\,MeV, and it
decays mainly into $\rm K^*K$. Hence the state is identified 
with the $\eta(1475)$.

The Crystal Barrel collaboration searched for the
$\eta (1295)$ and $\eta (1440)$ in
the reaction $p\bar p\to\pi^+\pi^-\eta (m)$,  
$\eta (m)\to\eta\pi^+\pi^-$\cite{Reinnarth} where
$m$ is a running mass. 
A clear pseudoscalar resonance signal was observed
at 1405\,MeV. 
A scan for an additional $0^+ 0^{- +}$ resonance 
provided no evidence for the $\eta (1295)$. A
resonance at 1480\,MeV was seen, with $M=1490\pm 15,
\Gamma=74\pm 10$. Again, there is no reason why the $\eta(1405)$
and the $\eta(1475)$ were observed but not the $\eta(1295)$.
The latter resonance is seen only in the charge exchange reaction
$\pi^- p\to n\eta\pi\pi$. In the 1970's, the $a_1(1260)$ properties 
were obscured by the so--called Deck effect
($\rho$--$\pi$ re-scattering in the final state). Possibly,
$a_0(980)\pi$ re-scattering fakes a resonance--like behavior; 
otherwise $\eta(1295)$ might be 
mimicked by feed--through from $f_1(1285)$. In any case, we
exclude  $\eta (1295)$ from the further discussion.
\par
The $\eta (1440)$ is  not
produced as $\bar ss$ state but decays with a large fraction
into $\rm K\bar K\pi$ and it is split into two components.
These anomalies are likely due
to a node in the $\eta(1440)$ wave function. 
The node has an impact on the decay matrix element which
were calculated by\cite{Barnes:1996ff} within the $^3P_0$ model.
\par
\begin{figure}[bht] 
\begin{tabular}{ccc} 
\hspace*{-6mm}\includegraphics[width=0.38\textwidth,height=5cm]{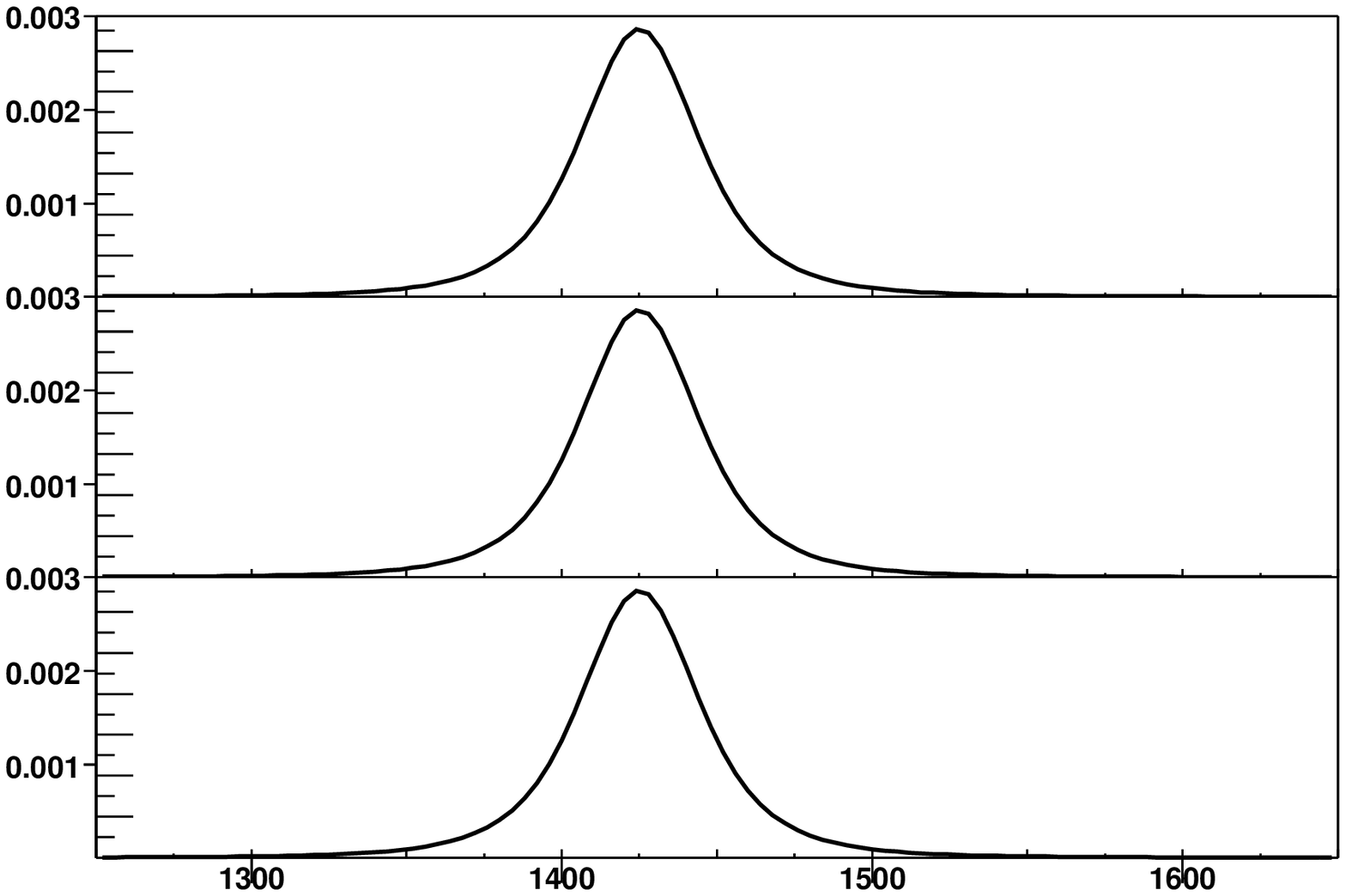}&
\hspace*{-9mm}\includegraphics[width=0.38\textwidth,height=5cm]{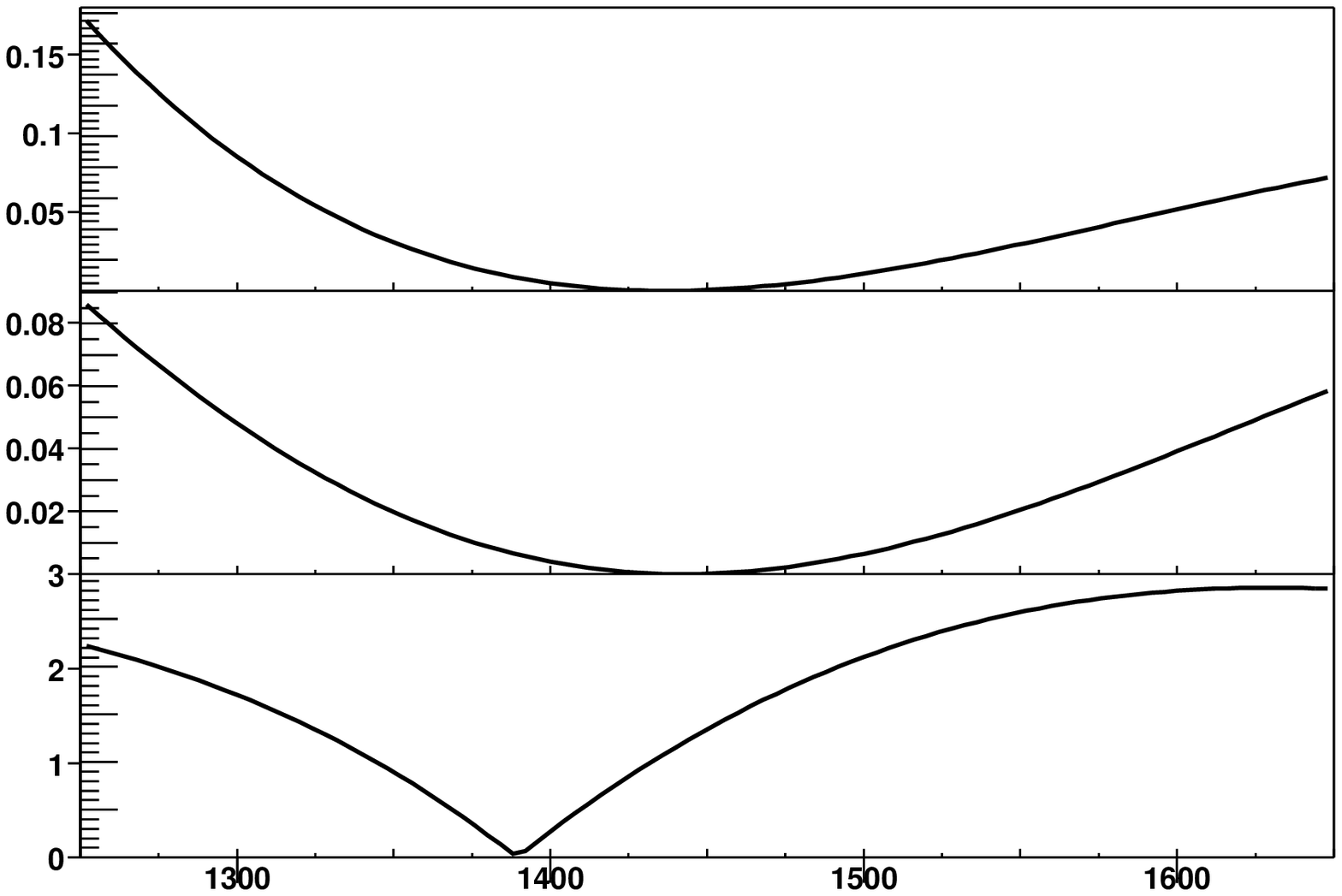}&
\hspace*{-8.5mm}\includegraphics[width=0.38\textwidth,height=5cm]{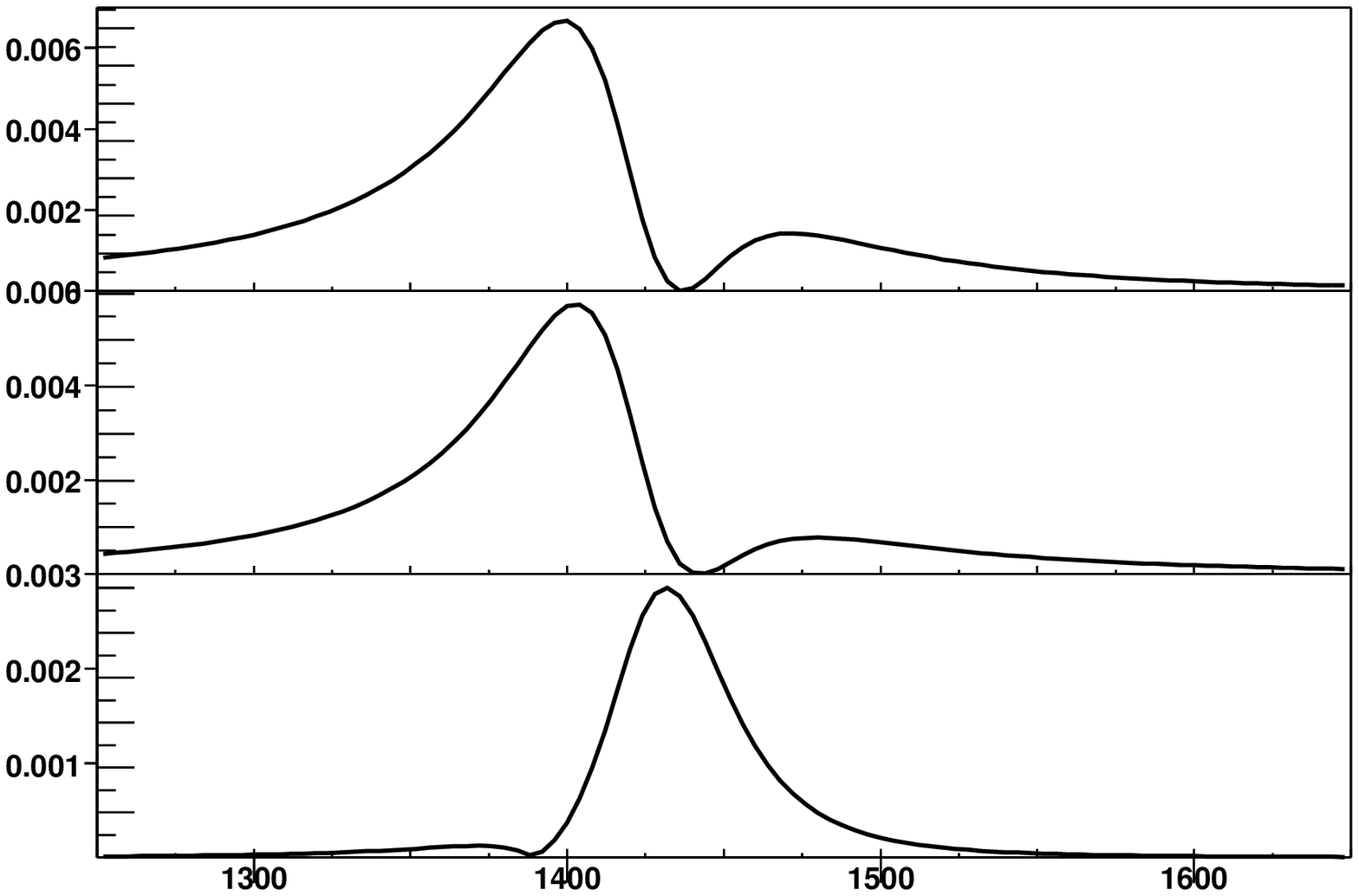}
\end{tabular}
\vspace*{-2mm}
\caption{\label{node}
Amplitudes for $\eta(1440)$ decays to $a_0\pi$ (first row), 
$\sigma\eta$ (second row):  $\rm K^*\bar K$ (third row)
the Breit-Wigner functions are shown on the left, then the squared 
decay amplitudes~\protect\cite{Barnes:1996ff} and, on the right, 
the resulting squared transition matrix element.
}
\end{figure}
The $\eta(1440)\to a_0(980)\pi$ and
$\to\rm K^*K$ distributions peak at different masses, 
about at 
the $\eta(1405)$ and $\eta(1475)$ masses. Hence there is no need to introduce
the $\eta(1405)$ and $\eta(1475)$ as two independent states. One
$\eta(1420)$ and the assumption that it is a radial excitation
describes the data. The phase motions 
of the
$a_0(980)\pi$ or $\sigma\eta$ isobar\cite{Reinnarth}
are compatible with only one pseudoscalar resonance being present in
the mass range from 1200 to 1500\,MeV. 
Hence the following states are identified as pseudoscalar
ground states and radial excitations:
\begin{center}
\renewcommand{\arraystretch}{1.1}
\begin{tabular}{lcccc}
\hline\hline
$1^1S_0$& $\pi$ & $\eta^{\prime}$ & $\eta$ & K  \\
$2^1S_0$& $\pi(1300)$ & $\eta(1760)$ & $\eta(1420)$ & K(1460) \\
\hline\hline
\end{tabular}
\renewcommand{\arraystretch}{1.0}
\end{center}

\subsection{Search for the scalar glueball}
The lowest--mass glueball has scalar quantum numbers. Its 
predicted mass ($\sim 1700$\,MeV)
falls into a region in which one may hope to get a consistent
picture of the mass spectrum of all scalar mesons.
Table~\ref{scalar-data} lists the spectrum of scalar
mesons as given by the Particle Data Group. 
\par
The $f_0(600)$, the lowest mass scalar meson, is 
often called $\sigma (600)$. The Particle Data Group
assigns to it a mass range from 400 to 1200\,MeV. In 
partial wave analyses, it is seen as a pole at about 500\,MeV.
However, the phase reaches $90^{\circ}$ only at $\sim 780$\,MeV.
A similar pole is observed in $\rm K\pi$ scattering; it is
often called $\kappa(800)$.
The nature of $\sigma$ and $\kappa$
is hotly debated: they may be $qq\bar q\bar q$ 
mesons\cite{Jaffe:1976ig}, 
relativistic S-wave $q\bar q$ states
(`chiralon')s\cite{Ishida:eg}, or
they might be due to attractive $\pi\pi$ or $\rm K\pi$ interactions, 
generated by `left--hand cuts' in a technical language.
Practically, the $\sigma(600)$ and 
$\kappa(900)$ do not play a role in the discussion of glueballs,
and the reader is referred to a recent review\cite{Tuan:2003bu}. 
The  $a_0(980)$ and $f_0(980)$ are often considered
as $\rm K\bar K$ molecular--like bound 
states\cite{Weinstein:gu,Janssen:1994wn,Locher:1997gr}. 
\par

\begin{table}[h!]
\renewcommand{\arraystretch}{1.1}
\caption{\label{scalar-data}The scalar mass
spectrum\protect\cite{Eidelman:2004wy}.}
{\begin{tabular}{cccccc}
\hline\hline
I = 1/2  & I = 1   &  I = 0    &         & \\  %
\hline
              &             & $\ \ f_0(600)$  &
&  ${\ \ \sigma (600)\ meson}$                  \\   
 {\ K(900)}        &             &             &        & \ 
{\ chiral partner of the $\ \pi$} \\  
              & $\ \ a_0(980)$  & $\ \ f_0(980)$  &        & $\ \ \rm K\bar K\ molecules$   \\
              &             &             &        & \\                 
\hline
              &             & $\ \ f_0(1370)$ &        &$\ \ q\bar q$ state\\  
$\ \ K_0^*(1430)$ & $\ \ a_0(1490)$ &
$\ \ f_0(1500)$ &        &2 $\ \ q\bar q$ 
{\ states}, {\ glueball} \\ 
              &             &             &        &  \\    
              &             & $\ \ f_0(1710)$ &        &$\ \ q\bar q$ state  \\    
\hline
$\ \ K_0^*(1950)$ &             &             &        &$\ \ q\bar q$ state  \\  
              &             & $\ \ f_0(2100)$ &
&$\ \ q\bar q$ state  \\  
              &             & $\ \ f_0(2200,2330)$ &   &$\ \ q\bar q$ state  \\  
\hline\hline
\end{tabular}
}
\renewcommand{\arraystretch}{1.0}
\vspace*{-3mm}
\end{table}
\par
The Crystal Barrel collaboration proposed the existence of two 
further scalar
isoscalar mesons, the $f_0(1370)$ and $f_0(1500)$. Their decays
were studied in a series of 
analyses\cite{Amsler:1995gf}$^-$\cite{Abele:1996nn}.
Three striking peaks  were observed in the
$\eta\eta$ invariant mass spectrum produced 
in $\bar pp$ annihilation in flight into $\pi^0\eta\eta$\cite{E760},
$1500, 1750$ and $2100$\,MeV. 
The data were not decomposed into partial waves
in a partial wave analysis, so the peaks could have 
J$^{\rm P\rm C}=0^{++}$, $2^{++}$, or higher. If the states had
J$^{\rm P\rm C}=2^{++}$, their decay into $\eta\eta$ would 
be suppressed by the angular momentum barrier. 
The peaks were seen very clearly suggesting 
$0^{++}$ quantum numbers. The same pattern of states 
was seen at BES in radiative J/$\psi$ decays\cite{bes4p}
into $2\pi^+ 2\pi^-$ . A partial wave analysis 
confirmed their scalar nature as had been suggested before in
a reanalysis of MARKIII data\cite{tokibugg}. 
The $f_0(1500), f_0(1710)$ and the $f_0(2100)$ have a similar 
production and decay pattern. Neither  $f_0(1370)$ nor 
`background' intensity was assigned to the scalar isoscalar 
partial wave. 
\par 
The first interpretation\cite{Amsler:1995td} of the scalar spectrum, 
also adopted by the Particle Data Group,
identifies the $a_0(980)$ and  $f_0(980)$ as non--$q\bar q$ states.
Then there are 10 states in the mass region of interest 
while the quark model predicts only 9
(3\,$a_(1450)$, 4\,K$^*_0(1430)$, and 2\,$f_0$'s). One of
the states,  $f_0(1370)$, $f_0(1500)$ or $f_0(1710)$,
must be the scalar glueball\,!
However, the  $f_0(1500)$ couples strongly to $\eta\eta^{\prime}$;
these are two SU(3) orthogonal states and cannot come from a singlet.
The  $f_0(1500)$ must hence have a strong flavor--octet component, 
it cannot be a pure glueball. The $f_0(1370)$ and $f_0(1500)$
decay strongly to $2\pi$ and into  $4\pi$ and weakly to
$\rm\bar KK$, they both cannot carry a large $\bar ss$ component.
The  $f_0(1370)$ is, probably, too light to be the scalar
glueball. So, none of the three states 'smells' like a glueball.
A way out is mixing; the two scalar $\bar qq$ states and the scalar
glueball have the same quantum numbers, they mix and 
form the three observed states. Table~\ref{scalar-data} summarizes
this interpretation. Several explicit mixing scenarios have been
suggested\cite{Amsler:1995td}$^-$\cite{Amsler:2002ey}
and some of them are capable of reproducing the decay
pattern. 
\par
An important ingredient of 
the `narrow--glueball' is the interpretation of the
$f_0(980)$ and $a_0(980)$ as alien objects, not related 
to $q\bar q$ spectroscopy. Several experiments
were directed to determine the structure of these two mesons,
like two-photon production\cite{Boglione:1999rw},
$\Phi$ radiative decay into 
$f_0(980)$\cite{Achasov:2000ym,Aloisio:2002bt}
and into $a_0(980)$\cite{Achasov:2000ku,Aloisio:2002bs},
and $Z^0$ fragmentation   \cite{opal}.
The conclusions drawn from these results are  
ambiguous.
Presumably, the wave function of the $f_0(980)$ and $a_0(980)$ 
is not just $\rm K\bar K$ but has a complex 
mass and momentum dependence. 
Likely, the outer part of the wave
function contains a large $\rm K\bar K$ component, in particular
close to the $\rm K\bar K$ threshold. The core however may be
dominantly $q\bar q$. 
In meson--meson scattering in relative S--wave, 
coupled channel effects play a decisive role. The
opening of thresholds attracts pole positions and the resonances
found experimentally do not agree with masses as
calculated in quark models. Under normal circumstances,
$K$--matrix poles, poles of the scattering matrix $T$ and
positions of observed peaks agree approximately, and the interpretation is
unambiguous. In S--waves, the situation is more complicated.
The mass of the resonance as quoted by experiments is the
$T$ matrix pole.
Quark models usually do not take into account the couplings
to the final state. Here, 
the $K$--matrix poles are compared to
quark model results, Table~\ref{aas}. The $K$--matrix poles come
from a series of coupled--channel
analyses\cite{Anisovich:1996qj,Anisovich:1997qp,Anisovich:2002ij},
mean values and errors are estimates provided by one
of the authors\cite{andrey}. 
The quark model states are from the Bonn model\cite{Koll:2000ke}, 
with the Lorentz structure B of the confinement potential.
Excellent agreement is observed. The two lowest scalar nonets are
identified, and there is one additional state, the 
$f_0(1400\pm 200)$. Its couplings to two pseudoscalar mesons
are flavor--blind, it is an isoscalar state. So it can be
identified as a scalar glueball. The width is problematic, 
it exceeds 2\,GeV. An excellent review of this approach can be found 
in\cite{Anisovich:1997qp}. 
The large scalar isoscalar background amplitude 
has been suggested as the scalar glueball  
in\cite{Morgan:td} and\cite{Minkowski:1998mf}.

\begin{table}
\renewcommand{\arraystretch}{1.1}
\caption{\label{aas}
The K--matrix poles of \protect\cite{andrey} show a remarkable
agreement with the results of the Bonn model~\protect\cite{Koll:2000ke}, 
version B. There is an
additional pole at $1400\pm200$\,MeV 
far off the real axis (i.e. $\sim 1000$\,MeV broad),
which is a flavor singlet and could be the glueball. 
}
{\begin{tabular}{ccc|ccc}
\multicolumn{3}{c}{K-matrix poles}&
\multicolumn{3}{c}{Bonn model, B}\\
\hline\hline
             &$a_0(980\pm30)$
             &$f_0(680\pm50)$ &
             &$a_0(1057)$
             &$f_0(665)$       \\  
             &&&&&\\  
             &           &           &        
             &           &                \\  
              $\rm K_0^*(1230\pm40)$
             &$a_0(1630\pm 40)$
             &$f_0(1260\pm 30)$
             &$\rm K_0^*(1187)$
             &$a_0(1665)$
             &$f_0(1262)$ \\  
             &&\hspace*{-2mm}$f_0(1400\pm 200)$
             &&&  \\ 
             &&$f_0(1600)$&
             &&$f_0(1554)$\\  
             &&&&&\\  
             $\rm K_0^*(1885^{+50}_{-100})$ &&&
             $\rm K_0^*(1788)$ &&\\
             &&$f_0(1810\pm 50)$
             &&&$f_0(1870)$     \\
\hline\hline
\end{tabular}
}
\renewcommand{\arraystretch}{1.0}
\end{table}
\par
Can the wide resonance be identified with a glueball\,?
This is neither known and nor tested. The ideal way to identify
the nature of such a broad state is a comparison of different
J/$\psi$ decay modes:
\begin{enumerate}
\item J/$\psi\to \omega\pi\pi$, \ J/$\psi\to \omega\rm K\bar K$, \ 
J/$\psi\to \omega\eta\eta$, \ J/$\psi\to \omega\eta\eta^{\prime}$, \
J/$\psi\to \omega 4\pi$ 
\item J/$\psi\to \phi\pi\pi$,  \ J/$\psi\to \phi\rm K\bar K$, \  
J/$\psi\to \phi\eta\eta$, \ J/$\psi\to \phi\eta\eta^{\prime}$, \
J/$\psi\to \phi 4\pi$ 
\item J/$\psi\to \gamma\pi\pi$, \ J/$\psi\to \gamma\rm K\bar K$,  \
J/$\psi\to \gamma\eta\eta$, \ J/$\psi\to \gamma\eta\eta^{\prime}$, \
J/$\psi\to \gamma 4\pi$ 
\end{enumerate}
\par
I anticipate that the data can be described by the pole positions 
given in Table~\ref{aas}. The glueball components of scalar mesons
do not couple to processes  (1) and (2) but only to (3). Thus the
glueball component can be identified.
Channels containing $\eta\eta$ and $4\pi^0$ would be the best choice
since a pion pair may also be produced from two primary gluons
by pion or $\rho$ exchange between the gluons, with
colour neutralization  by soft--gluon exchange. 
For $\eta\eta$ and $4\pi^0$ this process cannot occur. 
But also data recoiling against $\pi\pi$ and
$\rm K\bar K$ should allow a sensitive search for glueball
components.

\section{Is there convincing evidence for hybrids\,?}
The status of $J^{PC}=1^{-+}$ exotic mesons has recently been
reviewed\cite{sucek}. The lowest--mass candidate, $\pi_1(1370)$, 
decays into $\pi\eta$ and must be a four--quark state due to symmetry
arguments. A plethora of further four--quark states is then expected,
making unrealistic the attempt to identify one of them as hybrid.
The N(1440)\cite{Capstick:1999qq}  and the
$\Lambda(1600)$\cite{Kisslinger:2003hk} were proposed to be hybrid
baryons, but these interpretations are not compelling. The
$\Theta^+(1540)$, however, is a strong candidate for the anti-decuplet
expected from chiral soliton model.

\section{Conclusions}
There seems to be much more evidence for instanton--induced
interactions in hadron spectroscopy than for one--gluon exchange. 
There is no compelling evidence for gluons as
constituent parts in spectroscopy. 
Finally, the of the $\Theta^+(1540)$ - discussed intensively
at this workshop - seem to be more easily interpreted 
in the chiral--soliton--model than in models based on a special
five--quark dynamic.

\end{document}